\documentstyle[11pt,aaspp4]{article}
\lefthead{Meyer et al.}
\righthead{Small-Scale ISM Structure Toward M15}
\begin{document}
\singlespace
\tighten

\title{A Na I Absorption Map of the Small-Scale Structure in the
Interstellar Gas Toward M15}

\author{David M. Meyer\altaffilmark{1} and J. T. Lauroesch\altaffilmark{1}}
\affil{Department of Physics and Astronomy, Northwestern University, 
Evanston, IL  60208}
\authoremail{davemeyer@nwu.edu, jtl@elvis.astro.nwu.edu}

\altaffiltext{1}{Visiting Astronomer, Kitt Peak National Observatory,
National Optical Astronomy Observatories, which is operated by the
Association of Universities for Research in Astronomy, Inc. under
cooperative agreement with the National Science Foundation.}

\begin{abstract}
Using the DensePak fiber optic array on the KPNO WIYN telescope,
we have obtained high S/N echelle spectra of the Na~I D wavelength
region toward the central 27$\arcsec$~x~43$\arcsec$ of the
globular cluster M15 at a spatial resolution of 4$\arcsec$.
The spectra exhibit significant interstellar Na I absorption at
LSR velocities of $+$3~km~s$^{-1}$ (LISM~component) and
$+$68~km~s$^{-1}$ (IVC~component).  Both components vary appreciably
in strength on these scales.  The derived Na~I column densities
differ by a factor of 4 across the LISM~absorption map and by a
factor of 16 across the IVC~map.  Assuming distances of 500~pc
and 1500~pc for the LISM and IVC clouds, these maps show evidence
of significant ISM structure down to the minimum scales of
2000~AU and 6000~AU probed in these absorbers.  The smallest-scale
$N$(Na~I) variations observed in the M15 LISM and IVC maps
are typically comparable to or higher than the values found
at similar scales in previous studies of interstellar Na~I
structure toward binary stars.  The physical implications of
the small and larger-scale Na~I features observed in the M15
maps are discussed in terms of variations in the H~I column
density as well as in the Na ionization equilibrium.
\end{abstract}


\section{Introduction}

The evidence for significant subparsec-scale structure in the diffuse
interstellar medium (ISM) has been accumulating recently through
measurements of H I 21 cm absorption toward high-velocity
pulsars (\cite{fra94}) and extended extragalactic radio sources
(\cite{fai98}) as well as optical observations of the
interstellar Na I D absorption toward globular clusters (\cite{bat95})
and binary stars (\cite{mey96,wat96}).  At the pulsar
($\approx$10 to 10$^2$ AU), binary ($\approx$10$^2$ to 10$^4$ AU), and
globular cluster ($\approx$10$^4$ to 10$^6$ AU) scales sampled, all of
these observations imply dense concentrations of atomic gas
($n_H\gtrsim$ 10$^3$ cm$^{-3}$) in otherwise diffuse sightlines.  The
apparent ubiquity of this structure should be accounted for
in any successful ISM model.  However, due
to their large overpressures with respect to the intercloud medium,
such small-scale condensations cannot be accomodated in any abundance
by the standard \cite{mck77} pressure equilibrium model for the ISM\@.
\cite{hei97} has proposed a geometric solution where this
apparent structure is due to filaments or sheets of lower density
gas aligned along a given sightline that produce significant column
density differences (and spuriously high inferred volume densities) over
transverse length scales as small as 30~AU.  In an approach that removes the
requirement of pressure equilibrium, \cite{elm97} proposes a fractal
ISM model driven by turbulence that produces
self-similar structure down to the smallest scales.

A chief limitation impacting the interpretation of the diffuse ISM
structures observed to date has been the rather poor small-scale sightline
coverage.  In particular, each binary sightline samples the structure at
only one scale along one direction.  The few globular cluster studies have
typically involved 10 to 15 stars and have sampled only relatively large
separations.  However, the bright extended cores of some globulars do
provide a background source suitable for mapping the absorption-line
structure of intervening gas at much higher spatial resolution,
in two dimensions and with full sampling.
With a core $V$-band surface brightness of 14.21~mag~arcsec$^{-2}$
(\cite{har96}) falling to about 18~mag~arcsec$^{-2}$ at a radius of
30$\arcsec$ (\cite{her85}), the best example of such a cluster is M15
($d$~$=$~10.4~$\pm$~0.8~kpc; $v_{LSR}$~$=$~$-$99~km~s$^{-1}$).
Spectra of selected stars in M15 have revealed significant
interstellar Na I absorption at $v_{LSR}$ $=$ $+$3 and $+$68
km s$^{-1}$ that varies in strength on scales ranging from
about 1$\arcmin$ to 15$\arcmin$ (\cite{leh99,pil98,ken98,lan90}).
In this {\it Letter}, we present a fully-sampled, two-dimensional map of
the Na~I absorption over the central 27$\arcsec$ x 43$\arcsec$ of M15 as
part of a new effort to probe the patterns of such variations
down to scales of a few arc seconds.

\section{Observations}

The M15 observations were obtained in 1998 August using the DensePak
fiber optic array and Bench spectrograph on the 3.5~m
WIYN\footnotemark[2] telescope at Kitt Peak National Observatory.
The DensePak array consists of 91 fibers bonded into a 7 x 13
rectangle that covers 27$\arcsec$ x 43$\arcsec$ of sky with
center-to-center fiber (3$\arcsec$ diameter) spacings of
4$\arcsec$ at the WIYN F/6.4 Nasmyth focus
(\cite{bar98}).  The observations
were conducted with the centermost fiber positioned
at the center of M15 (RA $=$ 21$^h$ 29$^m$ 58.3$^s$,
Dec $=$ $+$12$\arcdeg$ 10$\arcmin$ 00$\arcsec$ (J2000.0)) and
the major axis of the array aligned along a N$-$S direction.
The spectrograph
was configured with the Bench camera, a Tek2048~CCD~(T2KC),
the Echelle grating, and an interference filter~(X17)
providing spectral coverage from 5725 to 5975~\AA\ at a 2.2~pixel
resolution of 0.27~\AA\ or 14~km~s$^{-1}$.

\footnotetext[2]{The WIYN Observatory is a joint facility of the
University of Wisconsin-Madison, Indiana University, Yale University,
and the National Optical Astronomy Observatories.}

Utilizing this instrumental setup in queue mode, a total of four
1300 s exposures were taken of M15 under sky conditions
characterized by $\approx$1$\arcsec$ seeing.  These raw CCD frames
were bias-corrected, sky-subtracted (using a 1300 s exposure of
adjacent blank sky), flat-fielded, combined, and
wavelength-calibrated using the NOAO IRAF data reduction package
to extract the net spectrum yielded by each fiber.  Based on
previous observations with the Bench spectrograph Na~I setup and on
data comparisons with the KPNO 4-m echelle spectrograph, the
uncertainty in the zero level of these spectra due to uncorrected
scattered light effects should be less than 3\%.  Accounting for
5 broken fibers and 3 others with low counts, 83 of the 91 fibers
produced usuable spectra with S/N ratios ranging from 30 at some
edgepoints of the array to over 150 nearer the center.  In order to
remove the telluric absorption in the vicinity of the Na~I
D$_2$~$\lambda$5889.951 and D$_1$~$\lambda$5895.924 lines, these
spectra were all divided by an atmospheric template based on
observations of several rapidly-rotating early-type stars with little
intervening interstellar matter.  Figure~1 displays the resulting
Na~I spectra for the center of M15 and three positions of various
separations and angles with respect to the center.  Three Na~I
doublets are apparent and well-separated in velocity
in all of these spectra $-$ the bluemost is due to stellar Na~I
absorption in M15, the middle or ``local ISM'' (LISM) component is
due to interstellar gas at $v_{LSR}$~$=$~$+$3~km~s$^{-1}$, and the
redmost or ``intermediate velocity'' (IVC) component is due to
presumably more distant gas at $v_{LSR}$~$=$~$+$68~km~s$^{-1}$.  It
is also apparent from Figure~1 that both the LISM and IVC Na~I
absorption toward M15 exhibit significant variations on scales
much less than~1$\arcmin$.  Over the whole map, the equivalent
width of the Na~I~D$_1$ line varies from 180 to 365~m\AA\ for
the LISM component and from 40 to 155~m\AA\ for the IVC component.

In terms of a surface map, Figures~2~and~3
show how the Na~I columns corresponding to the LISM
and IVC components vary across the face of M15 at the 4$\arcsec$
fiber resolution.  These column densities were measured using the
FITS6p profile-fitting package (\cite{wel94}) to simultaneously
fit the D$_2$ and D$_1$ lines in each fiber, assuming
single-component Voigt profiles for both the LISM and IVC
Na~I doublets.  Based on the higher
resolution ($\Delta$$v$~$=$~9.8~km~s$^{-1}$) Na~I data of
Kennedy et al.\ (1998) for two stars in M15, this assumption
should be reasonable for estimating the IVC column density but is
definitely an approximation to the multicomponent LISM absorption
structure.  In the case of the IVC fits, the derived
velocities exhibit a marginal increase of $\approx$1~km~s$^{-1}$
from N to S across the map and the derived line widths ($b$-values)
are typically near 3~km~s$^{-1}$ with some as low as 2.2~km~s$^{-1}$.
The lowest IVC Na~I column densities in Figure~3 correspond to
the weakest lines and have formal profile-fitting uncertainties of
about 10-20\%.  The highest IVC Na~I columns are more uncertain but
should be accurate to within a factor of two unless the profiles are
dominated by unresolved saturated structure that would mask even higher
columns.  In the case of the LISM fits, the derived $b$-values
(typically near 8~km~s$^{-1}$) lead to Na~I columns in Figure~2
that are probably underestimates of the ``true'' multicomponent
values but that are illustrative of the net equivalent width variations.

The only potential sources of stellar
contamination in measuring the LISM and IVC Na~I absorption
are the Ni~I~$\lambda$5892.883 and Ti~I~$\lambda$5899.304 lines
that would be located on the blue wing of the IVC~D$_2$
feature and near the center of the IVC~D$_1$~line, respectively.
Based on the F3~composite spectral type and low metallicity
([Fe/H]~$=$~$-$2.17) of M15 as well as the weakness of the
occasional excess absorption observed on the blue wing of the
IVC~D$_2$~line (which is mostly excluded by the single-component
fit), the impact of any stellar contamination on our
derived column densities is likely to be appreciably less
than the fitting uncertainties (\cite{mon99}).
The appearance of coherent
structures in Figures~2~and~3 suggests that the uncertainties
due to stellar line contamination are indeed smaller than
those due to the profile fitting.

\section{Discussion}

In order to discuss the ISM structure observed toward M15 in terms
of its physical length scales, it is necessary to estimate the
distances to the LISM and IVC clouds.  Given that \cite{alb93}
have measured weak Ca~II absorption near $v_{LSR}$~$=$~0~km~s$^{-1}$
toward HD~204862 ($d$~$\approx$~100~pc; 0.3$\arcdeg$ separation
from M15) and a much stronger line toward HD~203699
($d$~$\approx$~500~pc; 2.5$\arcdeg$ separation), we will assume
a distance of 500~pc for the M15 LISM absorption that should
at least be an upper limit for the
clouds comprising this column.  In the case of the IVC
component, Na~I absorption has been seen at a similar velocity
toward HD~203664 whose distance is about
3.2~kpc and angular separation from M15 is 3.1$\arcdeg$
(\cite{lit94,sem95,rya96}).  We will assume a distance of
1500~pc for the IVC absorber which should be accurate to
within a factor of two.  At these distances, the
27$\arcsec$~x~43$\arcsec$ coverage of the DensePak array corresponds
to a 13,500~x~21,500~AU (0.065~x~0.10~pc) section of the LISM
clouds and a 40,500~x~64,500~AU (0.20~x~0.31~pc) portion of the
IVC cloud.  The 4$\arcsec$~fiber spacing translates to spatial
resolutions of 2000 and 6000~AU for the LISM and IVC absorbers,
respectively.  The Na~I column densities are typically higher
in the LISM clouds with individual fiber values ranging from
2.3~x~10$^{12}$ to 8.5~x~10$^{12}$~cm$^{-2}$.  Over the minimum
2000~AU scale, the maximum $N$(Na~I) variation observed is
3.0~x~10$^{12}$~cm$^{-2}$ and the median $|$$\Delta$$N$(Na~I)$|$ is
4.5~x~10$^{11}$~cm$^{-2}$.  In the case of the IVC cloud, the
dynamic range in $N$(Na~I) is greater with values stretching
from 5.0~x~10$^{11}$ to 8.0~x~10$^{12}$~cm$^{-2}$.  The
maximum $N$(Na~I) variation observed over the minimum 6000~AU
scale in this cloud is 5.9~x~10$^{12}$~cm$^{-2}$ and the
median $|$$\Delta$$N$(Na~I)$|$ is 3.0~x~10$^{11}$~cm$^{-2}$.

Since the minimum scales probed by our M15 observations
overlap with those involving studies of
interstellar Na~I toward binary stars, it is important to
compare their evidence for small-scale ISM structure.  The
binary studies involve about 20 early-type systems
with projected linear separations mostly between 500 and
7000~AU and typical distances within several hundred pc
(\cite{mey96,wat96}).  Based on high-resolution
($\Delta$$v$~$\lesssim$~1.5~km~s$^{-1}$) Na I observations,
they find variations in at least one velocity component
toward each system that are collectively indicative of
ubiquitous small-scale structure.  The M15 observations
probe a sightline that is much longer and has a larger Na~I column
than most of these binaries at appreciably lower velocity resolution.
Also, whereas each star in a binary provides a single extremely
narrow beam ($\approx$0.0001$\arcsec$) through the intervening
ISM, each DensePak fiber samples a number of such beams from
the closely-spaced stars in the core of M15 and thus should smear out
any imprint of structure on scales appreciably smaller than the
3$\arcsec$~fiber diameter.  Nevertheless, the smallest-scale
$N$(Na I) variations observed in the M15 LISM and IVC
clouds are typically comparable to or larger than the values toward
the binary stars.  For example, the integrated column density
difference of 2.0~x~10$^{11}$~cm$^{-2}$ across the Na~I profile
toward the binary $\mu$ Cru (6600 AU separation) (\cite{mey96})
is similar to the median $|$$\Delta$$N$(Na~I)$|$ of
3.0~x~10$^{11}$~cm$^{-2}$ measured at the 6000~AU resolution
of the M15 IVC observations.  Since a circumstellar explanation
cannot be completely ruled out for at least some of the binary
Na~I variations, the M15 observations are important in providing
clear evidence of significant ISM structure on comparably
small scales where there is absolutely no possibility of
circumstellar contamination.

The key distinction between these M15 observations and
the binary studies is that here we image two ``clouds'' in two
spatial dimensions at a variety of scales whereas each binary
probe provides only a single measurement of a single scale for each
intervening cloud.  In comparing the characteristics of the LISM
and IVC maps, the texture of the LISM structure
appears to be generally smoother with larger angular features than in
the IVC gas.  This difference could be due to the greater IVC distance
or to the superposition of structures at different
distances within the LISM absorption profile.
An interesting feature of the LISM
map is how $N$(Na~I) goes from a relatively constant value along
the entire 21,500~AU length of the western edge to a generally
50\% higher value as one moves about 5000~AU to the east.
A 5000~AU separation binary oriented N$-$S would not be very sensitive
to this feature whereas an E$-$W orientation would yield a strong
signal of small-scale structure.
The most striking aspect of the IVC map is in the southern region
where the Na~I column density dives from 8.0~x~10$^{12}$~cm$^{-2}$
on the WSW edge to 1.3~x~10$^{12}$~cm$^{-2}$ and then back up to
5.6~x~10$^{12}$~cm$^{-2}$ on the ESE edge over a total
straight-line distance of 41,000 AU\@.  This behavior is more
suggestive of a clumpy structure with characteristic scales of
$\approx$10,000~AU and peak Na~I column densities that can rise
at least 5~times above the adjacent background.

Unfortunately, since Na~I is generally not a dominant ion in H~I clouds,
the physical interpretation of the Na~I structure apparent in 
Figures~2~and~3 is not clear.  However, studies of Galactic diffuse
clouds have shown that when $N$(Na I)~$\gtrsim$~10$^{12}$~cm$^{-2}$,
empirical relationships
can be utilized to estimate $N$(H) from $N$(Na~I) to within a
factor of two (\cite{hob74,sto78,wel94}).  Applying these
relationships to the significant southern clumps in the M15 IVC map
results in $N$(H)~$\approx$~5~x~10$^{20}$~cm$^{-2}$ and
$n_H$~$\approx$~1000~cm$^{-3}$ (assuming a roughly spherical geometry).
Interestingly, Kennedy~et~al.\ (1998) have mapped the H~I~21~cm
emission of the IVC cloud in the vicinity of M15 at 12$\arcmin$
resolution and found significant clumpy structure on these scales
with the highest column density ($N$(H~I)~$=$~4~x~10$^{19}$~cm$^{-2}$)
centered on the cluster and a quick dropoff
($N$(H~I)~$<$~10$^{19}$~cm$^{-2}$) on a 0.5$\arcdeg$ scale.  The fact
that the peak IVC H column inferred from the Na~I data is about 10
times greater than that from the 21~cm observations implies that
either there is significant H~I clumpiness within the radio beam
or the $N$(Na~I)/$N$(H) ratio in the IVC can be significantly higher
than that typically observed in the diffuse ISM\@.  In the case of the
former, this result would have important implications for
determining the metallicities of such halo clouds (both IVCs and
their higher-velocity HVC brethren).  Metallicities of $\approx$25\%
and $\approx$10\% solar have recently been derived for two HVCs
by comparing UV absorption measures of their S~II abundances toward
background quasars with much broader ($\ga$1$\arcmin$) 21~cm emission
beam measures of the intervening HVC H~I columns (\cite{lu98,wak99}).
If HVCs generally exhibit H~I structure of the magnitude and scale
implied by the M15 IVC Na~I data, these metallicities could seriously
be in error.  The metallicity question is important to resolve because
it has a direct bearing on the interpretation of HVCs as primarily
Galactic in origin through fountain phenomena (\cite{sha76,bre80}) or
as infalling lower metallicity extragalactic matter (\cite{bli99}).

At the same time, it is possible that $N$(Na~I)/$N$(H) rather than
$N$(H) is varying on small scales in the M15 IVC cloud.
\cite{lau98} have discovered that the $N$(Na~I)
differences observed toward the binary $\mu$~Cru are not seen
in the dominant ion Zn~II (which should mirror variations in $N$(H)).
They suggest that these differences are due to small-scale
variations in the Na~ionization equilibrium that are driven by
temperature and/or electron density fluctuations.  Although
the most significant $N$(Na~I) variations in the M15~IVC and LISM
maps are appreciably greater than those toward $\mu$~Cru, it is
important to note that small fluctuations in $n_H$
can amplify $N$(Na~I) since the Na~I column should
scale roughly as ${n_H}^2$ if $n_H$ is proportional to $n_e$
(\cite{peq86}).  For example, the highest $N$(Na~I) peaks in the IVC
map could be produced by increasing $n_H$ by a factor of $\approx$2.3
over adjacent background without any change in $N$(H~I).  However,
if $n_e$/$n_H$ is not a constant (as might be expected if partial H
ionization augments the electron supply from C photoionization), the
IVC $N$(Na~I) variations could be less reflective of $n_H$ and
more indicative of appreciable small-scale $n_e$ fluctuations.
Of course, it is not clear how such fluctuations could occur in a
cloud of low extinction far from any ionizing source.

In summary, our observations show that the LISM and IVC~gas toward
M15 exhibits significant structure in terms of its physical conditions
and/or H~I~column density down to arc~second scales.
Although our sky coverage is too limited to analyze the observed
patterns in detail over their full extent, it does appear that
the Na~I data rule out both a very flat distribution on the
27$\arcsec$~x~43$\arcsec$ scale of the DensePak array and a random
distribution on the 4$\arcsec$ scale of the individual fibers.
Through further interstellar absorption-line mapping of M15 and other
globulars with fiber arrays like DensePak, it will
be possible to increase this sky coverage and characterize the
spatial structure of diffuse clouds in the Galactic disk and halo
over a large range of physically-interesting scales
that are difficult to probe otherwise.

\acknowledgments

It is a pleasure to thank Di Harmer, Daryl Willmarth, and the rest
of the KPNO WIYN queue observing team for obtaining the data.  Comments
by Dan Welty were very helpful in substantially improving the paper.
We would also like to acknowledge useful conversations with Chris
Blades, Ed Jenkins, and Caty Pilachowski.

\clearpage

\smallskip\smallskip
\begin{figure}
\plotfiddle{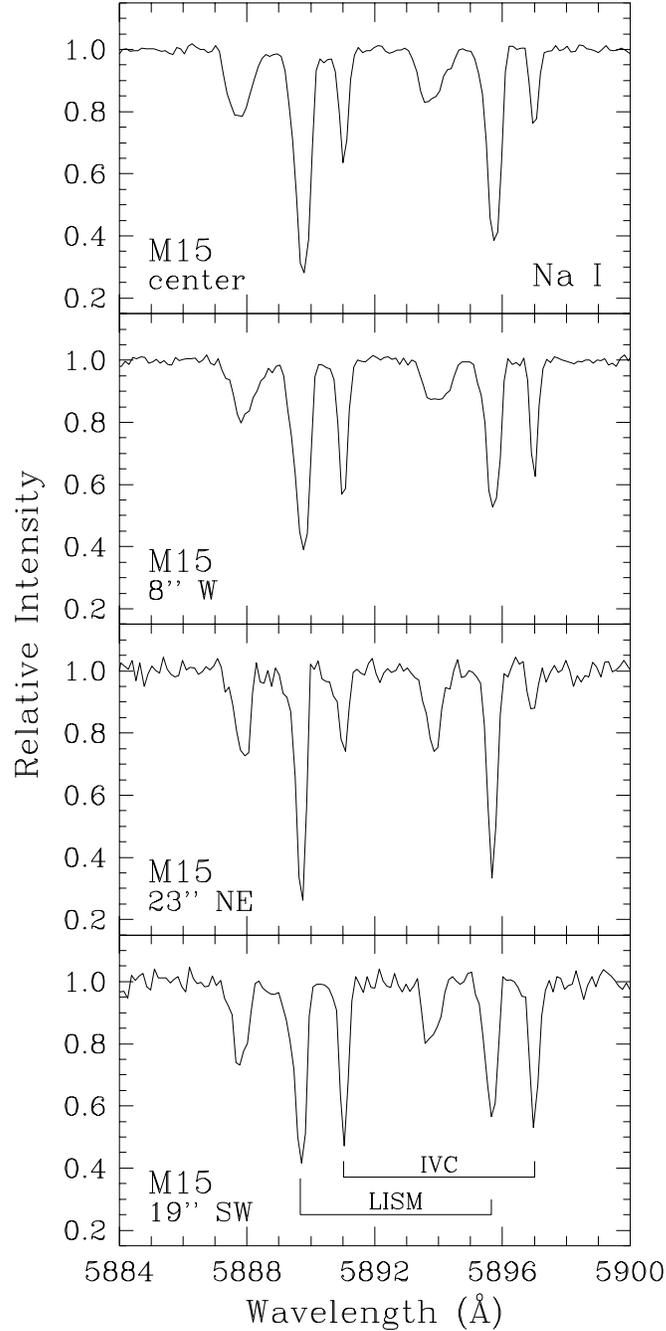}{6.8in}{0}{100}{100}{-324}{-216}
\caption[m1.ps]{WIYN DensePak echelle spectra of the Na~I
D$_2$ $\lambda$5889.951 and D$_1$ $\lambda$5895.924 region toward
the center of M15 and three labeled positions of various
separations and angles relative to the center.  The spectra have
a velocity resolution of 14~km~s$^{-1}$ and are displayed on
a heliocentric wavelength scale.  Among the three Na~I doublets
that are apparent in these spectra, the bluemost is due to
stellar absorption in M15, the component labeled ``LISM'' is
due to local interstellar gas at $v_{LSR}$~$=$~$+$3~km~s$^{-1}$,
and the component labeled ``IVC'' is due to presumably more
distant gas at $v_{LSR}$~$=$~$+$68~km~s$^{-1}$.}
\end{figure}

\begin{figure}
\plotone{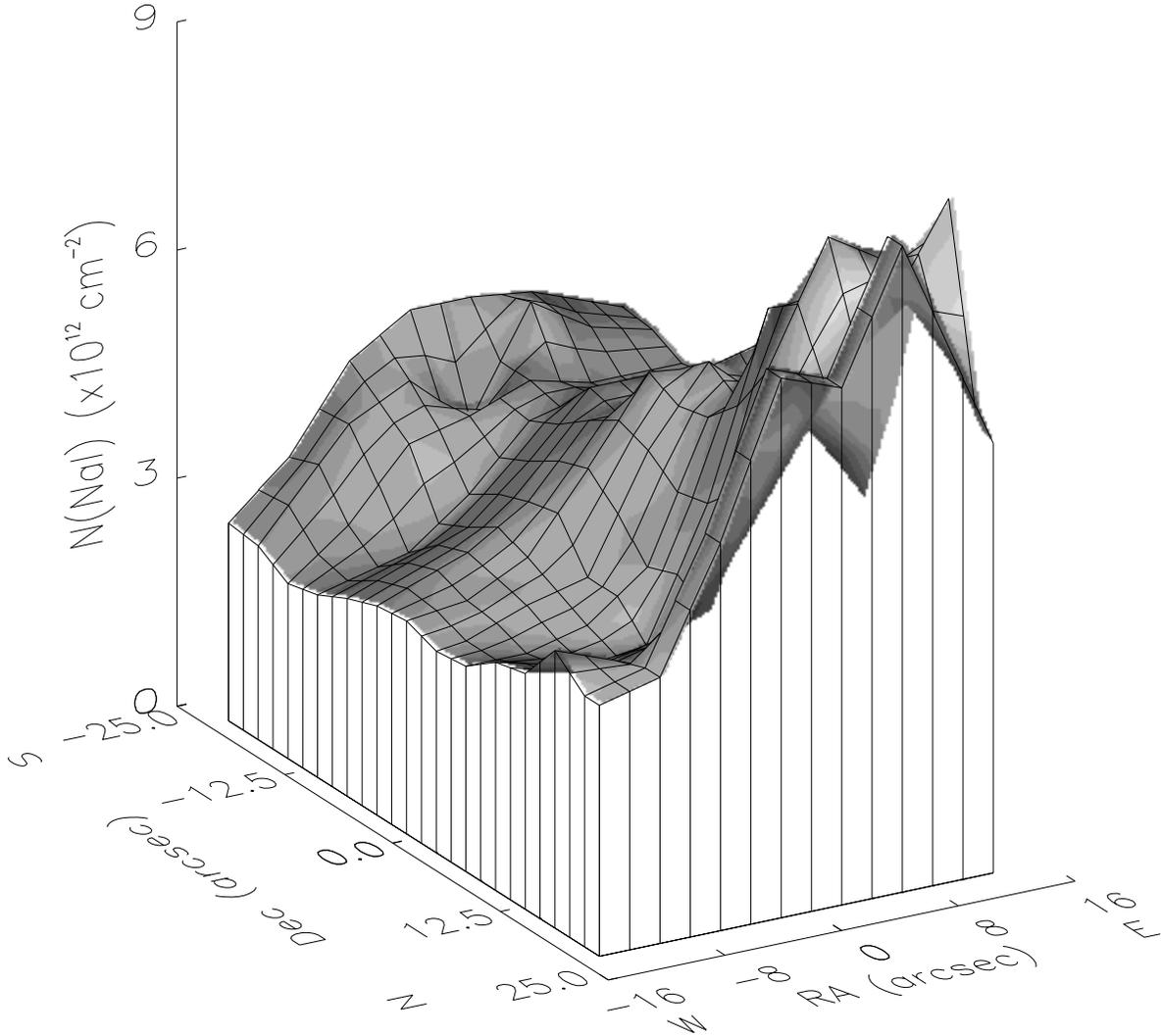}
\caption[m2r.ps]{A surface map of the Na~I column densities
corresponding to the LISM velocity component toward M15.  The
zero-point coordinates of the map refer to the center of M15
(RA~$=$~21$^h$~29$^m$~58.3$^s$,
Dec~$=$~$+$12$\arcdeg$~10$\arcmin$~00$\arcsec$~(J2000.0)).
The Na~I column densities were derived from individual fiber
spectra such as those displayed in Figure~1 using a
single-component Voigt profile fit.  Since the centers of
alternating rows of the DensePak array are offset by a
half-fiber in a honeycomb configuration, the column densities
were put into a 14~x~13 array by interpolating between points
in RA\@.  This surface plot was generated from a rebinning
of this array and has a spatial resolution of about 4$\arcsec$.
At the assumed distance of 500~pc for the LISM clouds, this
resolution projects to 2000~AU and the map covers
13,500~x~21,500~AU\@.}
\end{figure}

\begin{figure}
\plotone{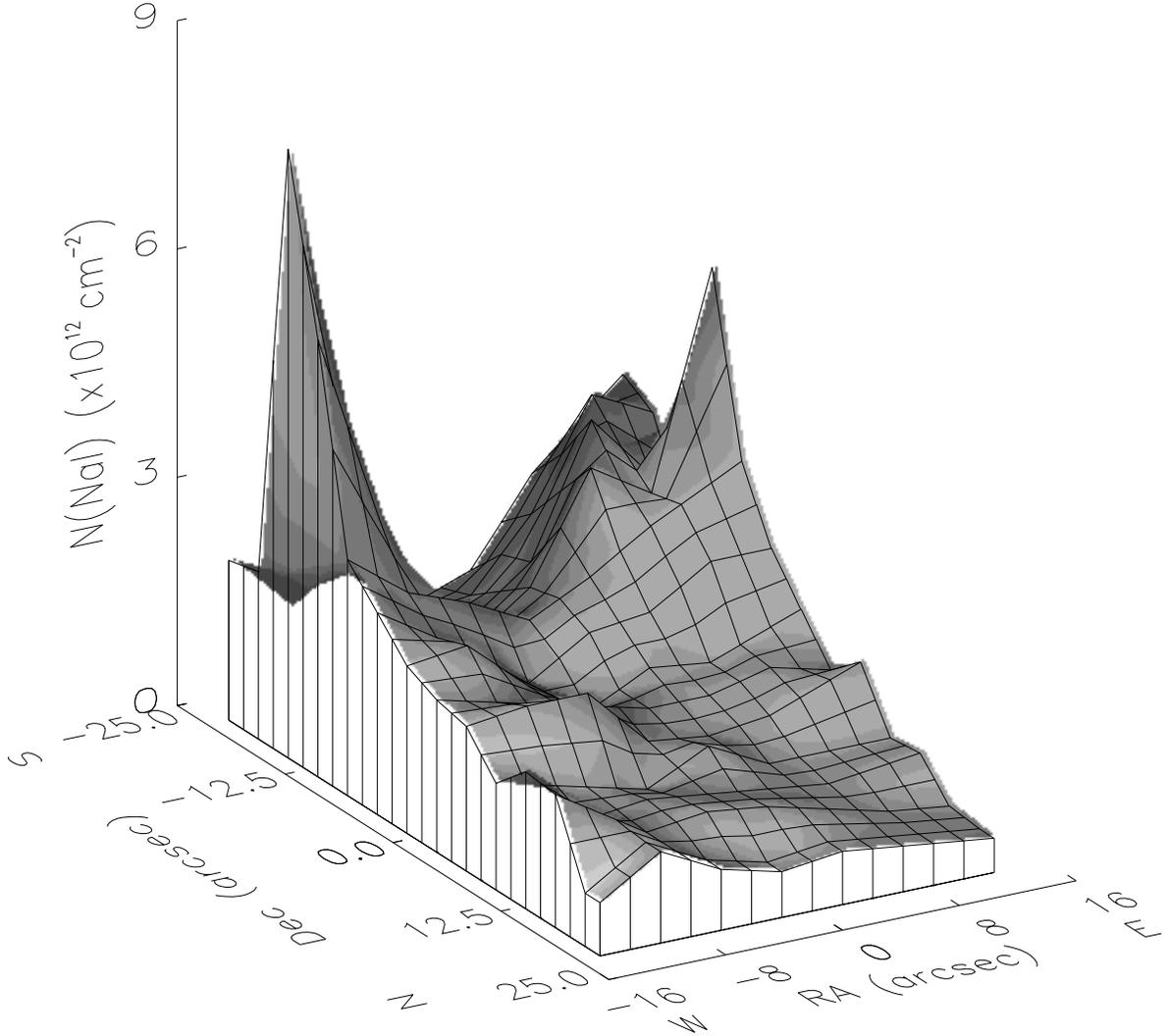}
\caption[m3.ps]{Same as Figure~2 for the IVC component
toward M15.  At the assumed distance of 1500~pc for the IVC
cloud, the 4$\arcsec$ resolution of the map projects to
6000~AU and the map covers 40,500~x~64,500~AU\@.}
\end{figure}

\end{document}